\documentclass[prb,preprint,showpacs,preprintnumbers,amsmath,amssymb]{revtex4-1}

\usepackage[dvipdfmx]{graphicx}
\usepackage{bm}

\newcommand{\be}{\begin{equation}}
\newcommand{\ee}{\end{equation}}
\newcommand{\eq}[1]{Eq.~(\ref{#1})}
\newcommand{\fig}[1]{Fig.~\ref{#1}}
\def\bea{\begin{eqnarray}}
\def\eea{\end{eqnarray}}

\def\bra{\langle}
\def\ket{\rangle}
\def\vq{{\bf q}}

\def\vk{{\bf k}}

\begin{document}

\title{Structure of the pairing gap from orbital nematic fluctuations} 

\author{Tomoaki Agatsuma$^{1,2}$ and Hiroyuki Yamase$^{1,2}$}
\affiliation{
{$^{1}$} Department of Condensed Matter Physics, Graduate School of Science, 
Hokkaido University, Sapporo 060-0810, Japan \\
{$^{2}$}National Institute for Materials Science, Tsukuba 305-0047, Japan
}

\date{\today}

\begin{abstract}
We study superconducting instability from orbital nematic fluctuations in a minimal 
model consisting of the $d_{xz}$ and $d_{yz}$  orbitals, and choose model parameters 
which capture the typical Fermi surface geometry observed in iron-based superconductors. 
We solve the Eliashberg equations down to low temperatures with keeping 
the renormalization function and a full momentum dependence of the pairing gap. 
When superconductivity occurs in the tetragonal phase, we find that 
the pairing gap exhibits a weak momentum dependence over the Fermi surfaces. 
The superconducting instability occurs also inside the nematic phase. 
When the $d_{xz}$ orbital is occupied more than the $d_{yz}$ orbital in the nematic phase, 
a larger (smaller) gap is realized on the Fermi-surface parts where 
the $d_{xz}$ ($d_{yz}$) orbital component is dominant, leading to 
a substantial momentum dependence of the pairing gap on the hole Fermi surfaces. 
On the other hand, 
the momentum dependence of the gap is weak on the electron Fermi surfaces. 
We also find that while the leading instability is the so-called $s_{++}$-wave symmetry, 
the second leading one is $d_{x^2-y^2}$-wave symmetry. 
In particular, these two states are nearly degenerate in the tetragonal phase 
whereas such quasi-degeneracy is lifted in the nematic phase and 
the $d_{x^2-y^2}$-wave symmetry changes to highly anisotropic $s$-wave symmetry. 
\end{abstract}

\pacs{74.20.Mn, 75.25.Dk, 74.20.Rp, 74.70.Xa}

\maketitle
\section{introduction}
The mechanism of high-$T_c$ superconductivity is one of major interests in condensed 
matter physics. In particular, iron-based superconductors (FeSCs) attract great interest \cite{kamihara08}. 
The typical phase diagram of FeSCs  (Ref.~\onlinecite{stewart11}) contains four phases: 
normal metallic phase, superconductivity (SC), spin-density-wave (SDW), and nematic phase \cite{fisher11}. 
Because of the proximity to the SDW phase, it is widely discussed that SC can be mediated by 
spin fluctuations \cite{mazin08,kuroki08,chubukov08}. 
On the other hand, FeSCs are characterized by multibands and thus 
SC mediated by orbital fluctuations is also discussed as another mechanism of 
SC (Refs.~\onlinecite{stanescu08} and \onlinecite{kontani11}). 

How about a role of the nematic phase for SC? 
Since SC occurs closer to the nematic than the SDW phase, 
it is easily expected that nematic fluctuations also play an important role to drive SC. 
While the nematic instability is accompanied by a structural phase transition 
from a tetragonal to an orthorhombic phase, 
the nematic phase is believed to be driven by electronic degrees of freedom, not by lattice degrees. 
Considering that the nematic phase is associated with breaking of 
the orientational symmetry and keeping the translational symmetry unbroken, 
strong nematic fluctuations are expected to occur around zero momentum near the nematic transition.  
In fact, such strong nematic fluctuations were directly observed by electronic Raman spectroscopy \cite{gallais13}. 
A possible SC from nematic fluctuations is therefore distinguished from 
the spin \cite{mazin08,kuroki08,chubukov08} and orbital  \cite{stanescu08, kontani11} fluctuation mechanisms 
because the latter two mechanisms are concerned with fluctuations of a large momentum transfer 
characterized typically by Fermi surface (FS) nesting.

The origin of the nematic phase is under debate \cite{fernandes14}. 
There are three possible nematic orders: charge \cite{kivelson98,yamase00,metzner00}, 
spin \cite{andreev84}, and orbital \cite{raghu09,wclee09} nematicity. 
The latter two possibilities, namely spin \cite{fang08,xu08}  and orbital \cite{krueger09,cclee09,lv09,baek15} 
nematic order, are mainly discussed. 
Since there is a linear coupling between spin and orbital nematic orders, 
one order necessarily leads to the other \cite{fernandes14}.  
It is therefore not easy to distinguish between these two orders in experiments. 
Theoretically it turned out that 
the spin nematic phase is subject to a severely restricted property near the SDW phase \cite{yamase15a}, 
which may serve to identify the origin of the nematic order. 

We focus on the orbital nematic scenario in this paper. 
Orbital nematic fluctuations lead to the so-called $s_{++}$-wave symmetry 
in the sense that it is $s$-wave and the gap has the same sign on all FSs 
(Ref.~\onlinecite{yanagi10b}). In the weak coupling limit 
without quasiparticle renormalization in the Eliashberg theory \cite{yanagi10a}, 
the transition temperature became unrealistically high and moreover 
the superconducting instability was restricted along 
the orbital nematic phase. These features were in sharp contrast to 
the typical phase diagram of FeSCs (Ref.~\onlinecite{stewart11}).  
Such drawbacks were overcome by taking 
quasiparticle renormalizations into account \cite{yamase13b}. 
The resulting onset temperature was decreased substantially down to a temperature 
comparable to experiments, suggesting that orbital nematic fluctuations 
can be a new mechanism driving high-$T_c$ SC. 
Furthermore orbital nematic fluctuations were found to drive strong coupling SC (Ref.~\onlinecite{yamase13b}). 
The pairing gap was, however, assumed to be constant on each FS and thus 
the structure of the gap, which is the fundamental property of SC, has not been clarified. 
 
In this paper, we study the momentum dependence of the pairing gap due to orbital 
nematic fluctuations by employing a minimal two-band model. 
We solve the Eliashberg equations down to low temperatures with keeping the renormalization function. 
We find that the momentum dependence of SC is very weak 
in the tetragonal phase whereas it becomes substantial on the hole FSs 
when SC occurs inside the nematic phase. These momentum dependences are 
understood in terms of multiorbital natures of SC. We also find that 
$d_{x^2-y^2}$-wave pairing is nearly degenerate to $s_{++}$-wave pairing 
when SC occurs from the tetragonal phase, whereas such quasi-degeneracy is 
lifted when SC occurs inside the nematic phase. 

In Sec. II we describe the model and formalism. Major results are presented in Sec. III 
and discussed in Sec. IV. Conclusions are given in Sec. V. 
In Appendix we present results deeply inside the nematic phase and 
the gap structure associated with subleading pairing instabilities.

\section{Model and Formalism}
To elucidate the typical feature of SC driven by orbital nematic fluctuations 
and to make feasible computations down to low temperatures, 
we employ a minimal model of orbital nematic physics. 
Since orbital nematic instability is described by the occupation difference between the $d_{xz}$ and 
$d_{yz}$  orbitals, we consider the following minimal interaction \cite{yamase13b} 
\be
H_{I} = \frac{g}{2} \sum_i n_{i-} n_{i-} \,. 
\label{HI}
\ee
Here $n_{i-}$ is the density-difference operator and is defined by
$n_{i-} = n_{i1}-n_{i2}$ with the electron density operator 
$n_{i\alpha} = \sum_\sigma c^\dagger_{i\alpha\sigma} c_{i\alpha\sigma}$. 
$i$ and $\sigma$ are site and spin indices, respectively, and $\alpha=1,2$
correspond to the $d_{xz}$ and $d_{yz}$  orbital, respectively. 
When the system retains the tetragonal symmetry, the expectation value of $n_{i-}$ 
becomes zero, namely $\bra n_{i-}\ket =0$. This expectation value becomes 
finite when the system loses $xy$ symmetry. 
Hence the quadratic form of $n_{i-}$ in \eq{HI} may be viewed as a typical interaction driving 
orbital nematicity. 
The coupling strength $g$ is an effective low-energy interaction 
coming from not only the bare intra-orbital Coulomb interactions \cite{misc-g}, 
but also the electron-phonon interaction \cite{yanagi10a},  
the Aslamazov-Larkin contribution \cite{kontani11}, 
and the interorbital Coulomb interaction between Fe and Pnictogen \cite{zhou11}. 
In principle, the interaction (\ref{HI}) can lead to a non-uniform solution of $\bra n_{i-} \ket$. 
However, in a parameter region we are interested in, the uniform solution 
gives the minimum energy in the random phase approximation (RPA). 

\begin{figure} [t]
\centering
\includegraphics[width=7cm]{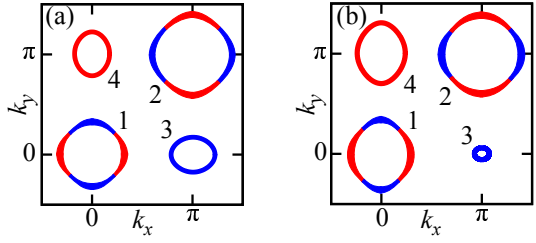}
\caption{(Color online)
Typical Fermi surfaces in the tetragonal phase (a) and the nematic phase (b).  
Fermi surfaces around $(0,0)$ and $(\pi,\pi)$ correspond to hole pockets and 
those around $(\pi,0)$ and $(0,\pi)$ electron pockets. 
Each FS is denoted as FS1, FS2, FS3, and FS4, respectively. 
Red and blue curves denote parts where the $d_{xz}$ and $d_{yz}$ orbital components 
are dominant, respectively, and the line width depicts its weight schematically.}  
\label{FS}
\end{figure}

The kinetic term of the two-band model may 
mimics the typical FSs in FeSCs (Refs.~\onlinecite{raghu08,yao09}): 
\be
H_0 = \sum_{{\bf k}, \sigma, \alpha,\beta}\epsilon^{\alpha\beta}_{\bf k}
c^\dagger_{{\bf k}\alpha\sigma} c_{{\bf k}\beta\sigma},
\label{H0}
\ee
where $\epsilon^{11}_{\bf k} = -2t_1 \cos k_x -2t_2 \cos k_y -4t_3 \cos k_x 
\cos k_y -\mu$, 
 $\epsilon^{22}_{\bf k} = -2t_2 \cos k_x -2t_1 \cos k_y -4t_3 \cos k_x 
\cos k_y - \mu$, and $\epsilon^{12}_{\bf k} = -4t_4 \sin k_x \sin k_y$. 
By choosing the parameters appropriate for FeSCs (Ref.~\onlinecite{yao09})  such as 
$t=-t_1$, $t_2/t=1.5$, $t_3/t=-1.2$, $t_4/t=-0.95$, and $\mu=0.6t$, 
we obtain two hole FSs around $\vk=(0,0)$ and $(\pi,\pi)$ 
and two electron FSs around $\vk=(\pi,0)$ and $(0,\pi)$ as shown in \fig{FS} (a). 
We denote them as FS$i$ with $i=1...4$. 
FS1 and FS2 are derived from both $d_{xz}$ and $d_{yz}$ orbitals and 
FS3 and FS4 are from the $d_{yz}$ and the $d_{xz}$ orbital, respectively. 
Our FSs capture the actual orbital components obtained in the 5-band model \cite{graser09}. 
Although the $d_{xy}$ orbital is partially involved in the electron FSs, 
the $d_{xy}$ orbital is not relevant to orbital nematicity and thus is neglected in the present model. 
For simplicity we use the unit cell containing one iron.

We  study the SC due to orbital nematic fluctuations in the framework of Eliashberg theory \cite{schrieffer}. 
We solve the Eliashberg equations down to low temperatures with keeping  the renormalization function 
as in the previous work \cite{yamase13b}. 
The key technical development of the present work is to include 
a full momentum dependence of the superconducting gap on each FS, 
which was neglected and replaced by a constant on each FS in the previous study \cite{yamase13b}. 

We compute nematic fluctuations in the RPA, which are expressed by 
$g(\vq, iq_m) = g \frac {\Pi_{0} (\vq, iq_m) }{1-g\Pi_{0}(\vq, iq_m)} g$, 
where $\vq$ and $i q_m$ are a momentum transfer and a bosonic Matsubara frequency, respectively.  
Here instantaneous contributions are subtracted to focus on the effect of nematic fluctuations. 
$\Pi_{0}(\vq, iq_m)$ describes a noninteracting nematic particle-hole excitations, 
namely $\Pi_{0}(\vq, i q_m) = \frac{T}{N} \sum_{\vk, \sigma, n} {\rm Tr} \left[ \mathcal{G}_{0} (\vk, i \omega_n) 
\tau_{3} \mathcal{G}_{0}(\vk+\vq, i \omega_n+ i q_m) \tau_{3} \right]$. 
Here $\mathcal{G}_{0}$ is a $2 \times 2$ matrix of the noninteracting Green function defined by  \eq{H0}, 
$\tau_{3}=\bigl( \begin{smallmatrix} 
1 & 0 \\ 0 & -1 
\end{smallmatrix} \bigr)$ 
is the vertex associated with the orbital nematic interaction [\eq{HI}], and $i\omega_n$ is a fermionic Matsubara 
frequency; $T$ is temperature and $N$ the total number of lattice sites. 

Since superconducting instability is a phenomenon close to the FS, we project the momenta on the FSs. 
We divide each FS into small patches and assign the Fermi momentum $\vk_{F}$ on each patch; 
$\vk_{F}$ is thus a discrete quantity in this paper.  
The resulting Eliashberg equations for the gap $\Delta(\vk_{F}, i\omega_n)$ and the renormalization
function $Z(\vk_{F}, i\omega_n)$ then read as 
\bea
&&\Delta(\vk_{F}, i\omega_n)Z(\vk_{F},  i\omega_n) = -\pi T \times
\nonumber
\\
&& \hspace{5mm} \sum_{\vk'_{F},n'}N_{\vk'_{F}}
\frac{\tilde{g}_{\vk_{F} \vk'_{F}} (i\omega_n-i\omega_{n'})}{|\omega_{n'}|}
\Delta(\vk'_{F}, i\omega_{n'})\,,
\label{eli1}
\\
&& Z(\vk_{F}, i\omega_n) = 1- 
\nonumber \\ 
&& \hspace{12mm}  \pi T\sum_{\vk'_{F},n'}N_{\vk'_{F}} \frac{\omega_{n'}}{\omega_n}
\frac{\tilde{g}_{\vk_{F} \vk'_{F}} (i\omega_n-i\omega_{n'})}{|\omega_{n'}|} \,.
\label{eli2}
\eea
Here $\tilde{g}$ denotes effective nematic fluctuations, which are obtained 
by averaging the nematic fluctuations over FS patches $\vk_{F}$ and $\vk'_{F}$. 
It is expressed by 
\bea
&&\tilde{g}_{\vk_{F} \vk'_{F}}  (i\omega_n-i\omega_{n'}) =  \nonumber \\ 
&& \hspace{2mm} \frac{
\frac{1}{N} \sum_{\vk}^{\rm FSp}  \frac{1}{N} \sum_{\vk'}^{\rm FSp}   
V(\vk, \vk')^{2} g(\vk-\vk', i\omega_n-i\omega_{n'})
}
{
\frac{1}{N} \sum_{\vk}^{\rm FSp}  \frac{1}{N} \sum_{\vk'}^{\rm FSp}   
} \,.
\label{gtilde} 
\eea
The sum over $\vk$ is limited to a FS patch specified by $\vk_{F}$. 
The vertex $V(\vk, \vk')$ describes a coupling between the nematic fluctuations 
and electrons, and is given by $V(\vk, \vk') = U^{\dagger}(\vk) \tau_3  U(\vk')$; 
$U$ is a $2 \times 2$ unitary matrix diagonalizing the kinetic term \eq{H0}. 
$N_{\vk_{F}}$ in Eqs.~(\ref{eli1}) and (\ref{eli2}) is a momentum resolved density of states defined on each FS patch. 
The renormalization function $Z(\vk_{F}, i\omega_n)$ is frequently neglected in research 
of FeSCs. However, its inclusion is definitely necessary because orbital nematic fluctuations 
lead to a strong coupling SC (Ref.~\onlinecite{yamase13b}). 
As is well known, \eq{eli1} can be viewed as an eigenvalue equation and 
the transition temperature $T_c$ is obtained when its eigenvalue $\lambda$ becomes unity.

\section{Results}
We study two typical cases, superconducting instability in the tetragonal phase ($g=-1.7t$)
and the nematic phase ($g=-1.8t$). 
The impact of nematic order on SC is also clarified. 

\begin{figure}[h]
\centering
\includegraphics[width=8.3cm]{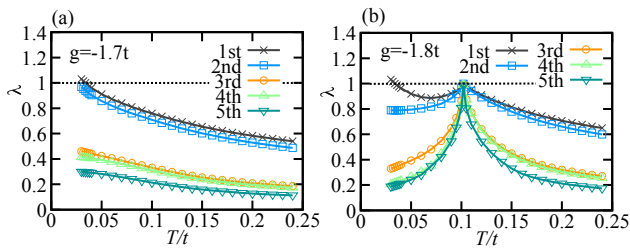}
\caption{(Color online)
Temperature dependence of the five largest eigenvalues $\lambda$ 
in the tetragonal phase (a) and the nematic phase (b). 
}
\label{lambda-T}
\end{figure}

\begin{figure}[ht]
\centering
\includegraphics[width=8.3cm]{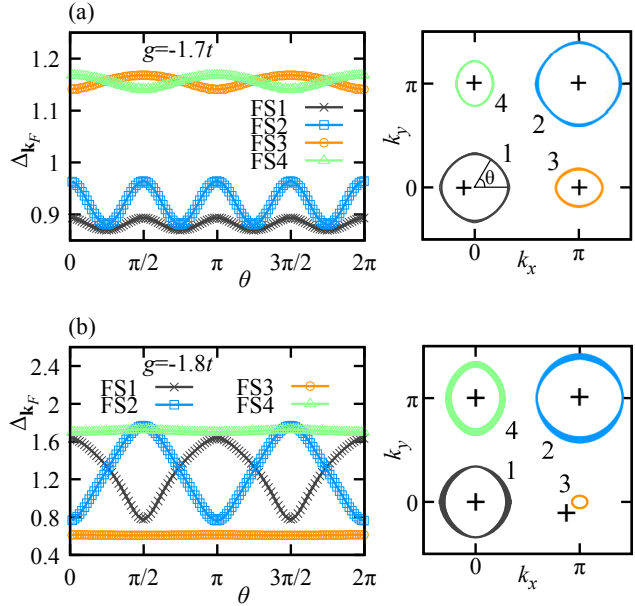}
\caption{(Color online)
Momentum dependence of the pairing gap on each FS in the tetragonal phase (a) 
and the nematic phase (b). 
In the right-hand panels, their momentum dependences are shown schematically by 
featuring a gap magnitude with the thickness of each FS. 
The polar angle $\theta$ is measured with respect to the $k_x$ axis for each FS. 
}
\label{gap}
\end{figure}

Figure~\ref{lambda-T}(a) shows the temperature dependence of the eigenvalue $\lambda$ 
for the five largest eigenvalues for $g=-1.7t$ where nematic instability does not occur 
down to zero temperature. 
With decreasing temperature, all $\lambda$s increase monotonically 
and the largest one eventually crosses unity at $T_{c}=0.034t$, where superconducting instability occurs. 
The corresponding eigenvector at the lowest Matsubara frequency, which we denote as 
$\Delta_{\vk_F} = \Delta (\vk_{F}, i \pi T_{c})$, shows $s_{++}$-wave symmetry 
as shown in \fig{gap}(a). 
The gap on FS1 and FS2 exhibits a fourfold modulation whereas 
the gap on FS3 and FS4 a twofold modulation even in the 
tetragonal phase.  
The modulation of the gap is very weak and is at most about 4\% on the hole FS (FS2). 
Comparison with the orbital components of the FSs in \fig{FS}(a) indicates that 
$\Delta_{\vk_{F}}$ on FS1 and FS2 is slightly suppressed on the FS parts 
where two orbital components contribute equally. 
A weak modulation of the pairing gap on the hole FSs is also obtained in the 
spin fluctuation mechanism \cite{maier09,thomale11,maiti11}, which however 
predicts a large modulation of the gap on the electron FSs, in contrast to the present 
orbital nematic mechanism.

The second largest eigenvalue is nearly degenerate to the leading $s_{++}$-wave gap in \fig{lambda-T}(a).   
It corresponds to $d_{x^2-y^2}$-wave symmetry where there are line nodes 
on FS1 and FS2 and a full gap on FS3 and FS4 with a sign opposite to each other. 
Interestingly a similar feature of such quasi-degeneracy of $s$- and $d$-wave solutions is obtained 
in the spin fluctuation mechanism \cite{kuroki08,graser09}. 
The third, fourth and fifth largest eigenvalues in \fig{lambda-T}(a) are rather suppressed.  
Details of their gap structure are presented in \fig{gap2-g1.7} in Appendix~B.

Figure~\ref{lambda-T}(b) shows the temperature dependence of eigenvalues for $g=-1.8t$. 
The  eigenvalues increase with decreasing $T$ and reaches close to $\lambda=1$ 
at the onset temperature of nematic instability 
$T_{\rm ON}=0.102t$. 
However, they do not cross unity. 
This is because the attractive interaction of $\tilde{g}_{\vk_{F} \vk'_{F}} (i\omega_{n}-i\omega_{n'})$ is 
strongly enhanced at low energy around $T=T_{\rm ON}$, but 
$Z(\vk_{F}, i\omega_n)$ also tends to diverge there, which then strongly reduces 
the quasiparticle weight and consequently suppresses superconducting instability. 
In $T<T_{\rm ON}$, the nematic order develops and thus low-energy nematic fluctuations are 
necessarily suppressed. 
Consequently the eigenvalues are also suppressed. However, the largest eigenvalue starts to grow 
again at lower temperatures, suggesting that orbital nematic fluctuations are still strong enough 
to drive SC. The largest eigenvalue eventually crosses unity at $T_c=0.034t$, 
leading to superconducting instability there. 
In contrast to the case of superconducting instability from the tetragonal phase [\fig{lambda-T}(a)], 
the second largest eigenvalue, which is characterized by nodal $s$-wave  symmetry 
[see \fig{gap2-g1.8}(a) in Appendix~B], 
is suppressed and no quasi-degeneracy 
of superconducting instability occurs in the nematic phase. 

While the orbital nematic order has two degenerate solutions, namely $\pm \bra n_{i -} \ket$, 
we consider a positive solution here. As a result, as shown in \fig{FS}(b), FS1 (FS2) 
elongates along the $k_y$ ($k_x$) direction, whereas 
FS4 expands along the $k_y$ direction and FS3 shrinks upon developing the nematic order. 
The corresponding eigenvector is plotted in \fig{gap}(b). In contrast to \fig{gap}(a), 
$\Delta_{\vk_{F}}$ shows a twofold modulation on FS1 and FS2.  
Its modulation amounts to as large as about 40\% with respect to its mean value. 
This strong modulation is understood in terms of the occupation difference of two orbitals in the nematic phase. 
In the present nematic phase, the $d_{xz}$ orbital is occupied more than the $d_{yz}$ orbital.   
Hence the contribution of the $d_{yz}$ orbital to the FSs becomes smaller than the other as seen in \fig{FS}(b). 
As a result, its contribution to the pairing are necessarily suppressed. 
Since FS1 and FS2 consist of both $d_{xz}$ and $d_{yz}$ orbitals [\fig{FS}(b)],  the pairing gap 
acquires substantial modulations on the hole FSs with minima where the $d_{yz}$ orbital is dominant 
as shown in \fig{gap}(b). 
It is interesting that the enhancement of the gap modulation in the nematic phase 
is also obtained in the spin fluctuation mechanism \cite{kang14}. 
On the other hand, FS3 and FS4 
consist of essentially a single orbital component and thus 
modulation of $\Delta_{\vk_{F}}$ retains very weak. 
The magnitude of the gap on FS3 becomes substantially smaller than that on FS4, because 
the minor $d_{yz}$ orbital  forms FS3.  

\begin{figure}[th]
\centering
\includegraphics[width=8.3cm]{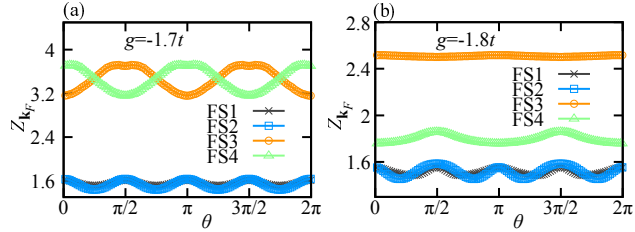}
\caption{(Color online)
Momentum dependence of the renormalization function $Z_{\vk_F}$ on each FS in the tetragonal phase (a) 
and the nematic phase (b). 
}
\label{Z}
\end{figure}

The momentum dependence of the renormalization function at the lowest Matsubara frequency 
is shown in \fig{Z}; here $Z_{\vk_F} = Z (\vk_{F}, i \pi T_{c})$. 
A value of $Z_{\vk_F} $ is substantially larger than the typical weak-coupling SC 
characterized by $Z_{\vk_F}$ close to unity. Hence 
orbital nematic fluctuations drive strong coupling SC (Ref.~\onlinecite{yamase13b}). 
On FS1 and FS2,  $Z_{\vk_F} $  shows the weak $\vk_{F}$ dependence 
in both tetragonal and nematic phases, 
with fourfold symmetry in the former and twofold symmetry in the later.  
On FS3 and FS4, $Z_{\vk_F} $ exhibits twofold modulation and its value is 
enhanced more than that on FS1 and FS2. 
This is because the size of the FS is rather small and thus orbital nematic fluctuations, 
which have large spectral weight at small momentum,  
contribute effectively via intra-FS scattering processes. 
In particular, the value of $Z_{\vk_F}$ 
amounts to as large as about 3.7 in the tetragonal phase. 

While we have considered the case where FS3 survives in the nematic phase, 
essentially the same results are obtained even if FS3 disappears due to large nematicity. 
Details are presented in Appendix~A. 

\section{Discussions}
The structure of the pairing gap can be revealed directly by angle-resolved photoemission spectroscopy (ARPES). 
We have obtained the weak momentum dependence of the gap in the tetragonal phase [\fig{gap}(a)],  
which can be viewed as a nearly isotropic gap. Such a gap roughly captures experimental observations 
in various materials when we focus on the FSs originating mainly from the $d_{xz}$ and $d_{yz}$ orbitals:  
BaFe$_{2}$(As$_{1-x}$P$_{x}$)$_{2}$ with $x=0.30$ \cite{yoshida14} and $0.35$ \cite{shimojima11}, 
Ba$_{0.6}$K$_{0.4}$Fe$_{2}$As$_{2}$ \cite{shimojima11}, 
FeTe$_{0.6}$Se$_{0.4}$ \cite{okazaki12}, and LiFeAs \cite{borisenko10,umezawa12}. 
For Ba(Fe$_{1-x}$Co$_{x}$)$_{2}$As$_{2}$ with $x=0.1$, a nearly isotropic gap 
was observed on all FSs except for one electron FS around $(\pi,0)$, where nodes or gap minima 
were reported \cite{hajiri14}. 
The presence of the node-like structure in the tetragonal phase 
is not captured in the present theory, which may be 
resolved by considering the following possibilities. 
First, while the $d_{xz}$ and $d_{yz}$ orbitals are dominant contributions 
to the Fermi level, the second dominant contribution comes from 
the $d_{xy}$ orbital \cite{graser09}. 
Since a modulation of the gap originates from the multiorbital natures in the present theory,  
a stronger modulation, namely gap minima, could be realized by 
including the $d_{xy}$ orbital. 
Second, the leading $s_{++}$-wave symmetry is nearly degenerate to 
the second leading instability [\fig{gap}(a)], which is characterized by $d_{x^2-y^2}$-wave symmetry.  
This $d_{x^2-y^2}$-wave symmetry could be stabilized by 
additional effects such as spin fluctuations. 
Third, spin fluctuations themselves, on the other hand, tend to drive 
$s_{\pm}$-wave symmetry in general \cite{mazin08,kuroki08,chubukov08} and 
their inclusion yields the competition with $s_{++}$-wave symmetry.  
Such competition may lead to a node-like feature of $s$-wave gap. 
This is indeed the case at least when the system contains both orbital fluctuations 
with large momentum transfers and spin fluctuations \cite{saito13}. 
These three possibilities may also apply to the understanding of nodal gaps in 
Ba$_{1-x}$K$_{x}$Fe$_{2}$As$_{2}$ with $x$ close to 1 (Ref.~\onlinecite{ota14}).

The superconducting gap structure was also revealed for FeSe films, whose $T_c$ can be more than 
65 K \cite{wang12,he13,ge15}.  
A nearly isotropic gap was observed on the electron FSs for monolayer FeSe \cite{Liu12,zhang16} and 
K-coated multilayer FeSe \cite{miyata15}. 
A similar gap structure was also observed for 
Cs$_{0.8}$Fe$_{2}$Se$_{2}$ \cite{zhang11a} and K$_{0.8}$Fe$_{2}$Se$_{2}$ \cite{zhang11a}. 
These results are consistent with our results [\fig{gap}(a)]. 
While FeSe films \cite{he13} and alkali-intercalated FeSe \cite{zhang11a} 
are special in the sense that hole FSs 
are absent and only the electron FSs exist, the present theory 
is expected not to be sensitive to the actual FS geometry (see also Appendix~A)  
since SC from orbital nematic fluctuations 
comes mainly from intra-FS scattering processes \cite{yamase13b}. 

A superconducting gap inside the nematic phase may be discussed for FeSe$_{0.93}$S$_{0.07}$ (Ref.~\onlinecite{xu16}). 
A gap on the hole FS around $(0,0)$ exhibits the sizable momentum dependence 
with gap maxima (minima) along the $k_x$ $(k_y)$ axis in Fig.~\ref{gap}(b), 
which captures the observed gap structure on the hole FS (Ref.~\onlinecite{xu16}). 
The SC gap on the hole FS around $(\pi,\pi)$ has maxima (minima) 
along the $k_{y}$ ($k_{x})$ direction as seen in Fig.~\ref{gap}(b). 
Recalling that FeSCs have two irons per unit cell, our Brillouin zone would be folded and thus 
the hole FS around $(\pi,\pi)$ is actually moved around $(0,0)$ 
through a momentum shift of $(\pi,\pi)$, forming the outer hole FS around $(0,0)$. 
Consequently, we expect an {\it antiphase} gap 
structure between two hole FSs around $(0,0)$, that is, the outer FS has a larger 
(smaller) gap along the $k_y$ ($k_x$) axis whereas the inner FS has 
a smaller (larger) gap there. This predicted gap structure as well as 
the gap on the electron FSs have not been resolved in experiments \cite{xu16}. 

\section{Conclusions}
Employing a minimal two-band model consisting of the $d_{xz}$ and $d_{yz}$ orbitals,  
we have studied typical properties of SC mediated by orbital nematic fluctuations.  
We have solved the Eliashberg equations down to the superconducting onset temperature 
with keeping not only the renormalization function but also a full momentum dependence 
of the paring gap on the FSs. We have found that the leading instability is 
$s_{++}$-wave symmetry. The pairing gap exhibits a fourfold and twofold modulation on the 
hole and electron FSs, respectively, in the tetragonal phase. 
The gap is suppressed on the parts of the FSs 
where two orbitals contribute equally, but its suppression is weak and the gap may be approximated as a constant. 
SC with $d_{x^2-y^2}$-wave symmetry can also be driven by orbital nematic fluctuations 
as a nearly degenerate state to the $s_{++}$-wave state. 
The impact of the nematic order is noticeable. 
First, the gap on the hole FSs acquires a significant modulation. 
The gap is suppressed on parts of the FSs where the $d_{yz}$ ($d_{xz}$) orbital becomes dominant, 
when the $d_{yz}$ ($d_{xz}$) orbital is occupied less than the other. 
Second, the fourfold modulation of the gap on the hole FSs changes to a twofold modulation 
whereas the twofold modulation retains on the electron FSs. 
Third, the quasi-degeneracy of $s_{++}$- and $d_{x^2-y^2}$-wave solutions 
is lifted in the nematic phase. 
The $d_{x^2-y^2}$-wave solution is suppressed by changing its symmetry to 
highly anisotropic $s$-wave state. 

We have focused on orbital nematic fluctuations in order to establish the typical 
gap structure of SC mediated by them, which will serve to disentangle 
complex phenomena with combined effects from multiorbitals and multifluctuations in FeSCs.  
Given that the nematic phase is realized close to the SDW phase in the general phase diagram of FeSCs, 
we consider it reasonable to assume that spin fluctuations are also important to SC. In fact, 
there are a plenty of studies trying to explain the superconducting gap in FeSCs 
in terms of the spin fluctuation mechanism \cite{hirschfeld16}. 
An important future issue is to clarify the condition of which mechanism, 
spin fluctuations or orbital nematic fluctuations, is dominant over the other   
or whether both mechanisms should be considered on an equal footing in general.  
Although these two mechanisms reply on different physics, interestingly they 
share some aspects of SC:  
i) the pairing gap with $s$-wave symmetry \cite{mazin08,kuroki08,chubukov08}, 
ii) the presence of a $d_{x^2-y^2}$-wave solution nearly degenerate to 
the leading instability in the tetragonal phase \cite{kuroki08,graser09}, 
iii) the weak modulation of the pairing gap on the hole FSs in the tetragonal phase \cite{maier09,thomale11,maiti11},   
and iv) its enhancement in the nematic phase \cite{kang14}.

\acknowledgments
The authors thank T. Shimojima for valuable discussions on unpublished data in his group 
and T. Terashima for many fruitful comments. 
This work was supported by a Grant-in-Aid for Scientific Research Grant Number JP15K05189.

\appendix
\section{Gap structure deeply inside the nematic phase} 

\begin{figure*}[tb]
\centering
\includegraphics[width=10cm]{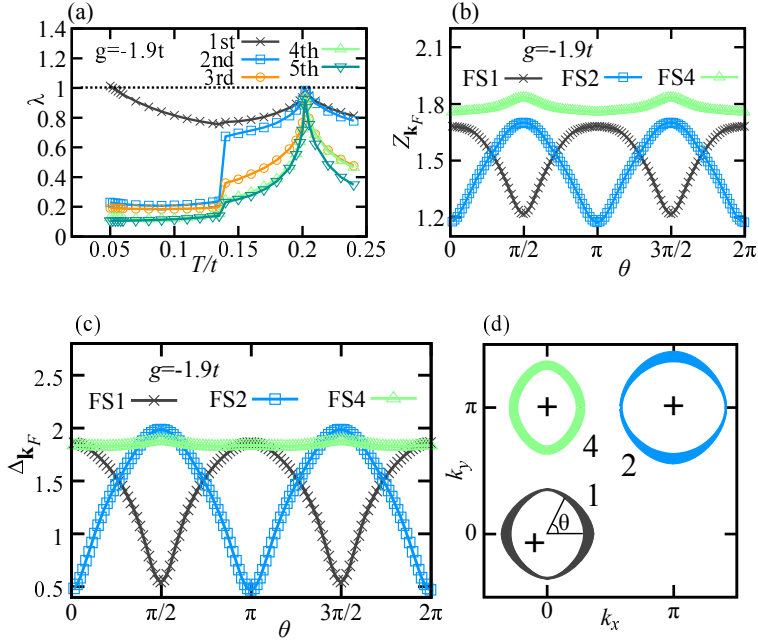}
\caption{(Color online) Major results deeply inside the nematic phase for $g=-1.9t$ 
where FS3 vanishes. 
(a) Temperature dependence of the five largest eigenvalues $\lambda$. 
(b) Momentum dependence of the renormalization function $Z_{\vk_F}$ on each FS. 
The polar angle $\theta$ is measured with respect to the $k_x$ axis for each FS 
as shown in (d). 
(c) Momentum dependence of the pairing gap on each FS. 
(d) Sketch of $\Delta_{\vk_{F}}$ by featuring a gap magnitude with the thickness of each FS. 
}
\label{g1.9}
\end{figure*}

Superconductivity mediated by orbital nematic fluctuations comes mainly from intra-pocket 
scattering processes \cite{yamase13b}. Hence the geometry of the FSs is not important to 
the superconducting instability. 
This is a crucial difference from other superconducting mechanisms such as 
spin fluctuations \cite{mazin08,kuroki08,chubukov08} and orbital fluctuations with a large momentum 
transfer \cite{stanescu08,kontani11}. To demonstrate this, we here present results of 
superconducting instability deeply inside the nematic phase ($g=-1.9t)$, 
where FS3 vanishes due to large nematicity and 
the other FSs, namely FS1, FS2, and FS4 are elongated slightly more than \fig{FS}(b),  
as seen in \fig{g1.9}(d).  

Figure~\ref{g1.9}(a) shows the temperature dependence of the eigenvalues. 
With decreasing temperature, the eigenvalues increase and take a cusp at $T=T_{\rm ON} =0.202t$, 
where the nematic instability sets in. 
The eigenvalues do not cross unity there because the quasiparticle  residue $Z^{-1}$ 
goes to zero there, as in the case of \fig{lambda-T}(b). 
Below $T_{\rm ON}$, low-energy orbital nematic fluctuations are suppressed, leading to 
the suppression of the eigenvalues. At $T \approx 0.14t$, the eigenvalues drop discontinuously. 
This temperature corresponds to the temperature at which FS3 vanishes because the nematic 
order parameter grows to be large enough to push up FS3 above the Fermi energy. 
The largest eigenvalue, however, starts to increase at lower temperature and 
finally leads to superconducting instability at $T_c=0.052t$. 

In \fig{g1.9}(c), we show the momentum dependence of the pairing gap. 
The results are essentially the same as \fig{gap}(b). 
A quantitative difference is that $\Delta_{\vk_{F}}$ acquires a larger modulation on the hole FSs. 
The regions on the FSs where the $d_{xz}$ and $d_{yz}$ orbital components are dominant 
are almost the same as \fig{FS}(b) except for the absence of FS3. 
The gap minima are then realized on the FS parts consisting mainly of 
the minority orbital, namely the $d_{yz}$ component. 
The resulting modulation of the pairing gap amounts to as large as about 60\%.  
In spite of the large modulations on FS1 and FS2, the gap on FS4 exhibits 
the very weak momentum dependence. 
This is because FS4 consists of essentially a single orbital (see \fig{FS}). 
We summarize the gap structure associated with subleading instabilities in \fig{gap2-g1.9} 
in Appendix~B.

The corresponding renormalization function is shown in \fig{g1.9}(b). 
In line with the large modulation of $\Delta_{\vk_{F}}$ on FS1 and FS2, 
$Z_{\vk_{F}}$ on FS1 and FS2 also shows a modulation as large as about 17\% 
with a twofold modulation much more clearly than the corresponding results in \fig{Z}(b) 
because of larger nematicity here. 
On the other hand, $Z_{\vk_{F}}$ on FS4 features the very weak momentum dependence, 
similar to that of $\Delta_{\vk_{F}}$.

\section{Momentum dependence of paring gap of subleading instabilities} 

We present the momentum dependence of the pairing gap associated with the second, third, fourth, 
and fifth largest eigenvalues shown in Figs.~\ref{lambda-T} and \ref{g1.9}(a). 

\begin{figure*}[ht]
\centering
\includegraphics[width=15cm]{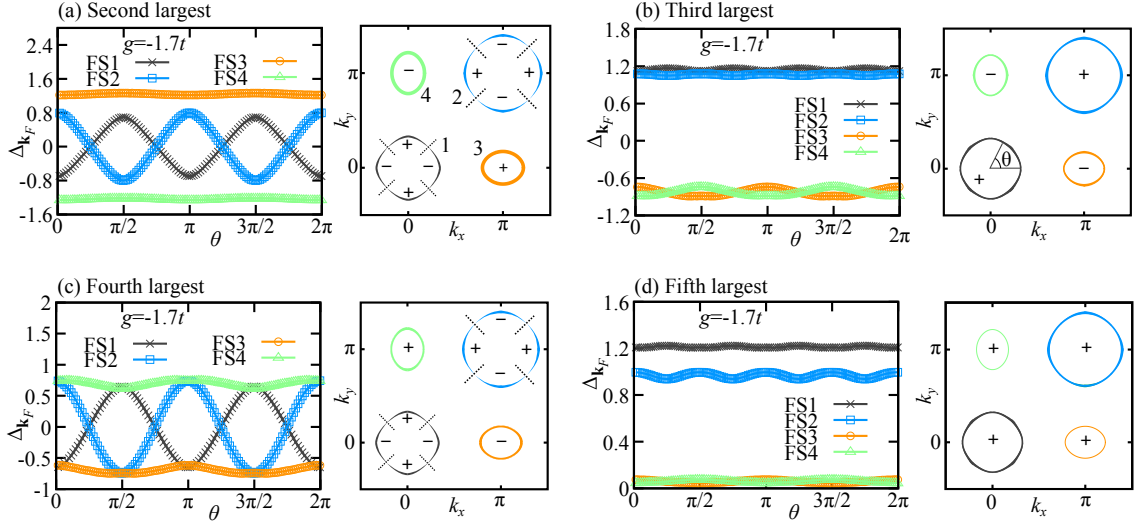}
\caption{(Color online) 
Momentum dependence of the pairing gap for the second (a), third (b), fourth (c), and fifth (d) 
largest eigenvalues in the tetragonal phase for $g=-1.7t$. 
The right-hand panel is a sketch of the gap structure by  
featuring a gap magnitude with the thickness of each FS. 
The polar angle $\theta$ is measured with respect to the $k_x$ axis for each FS. 
}
\label{gap2-g1.7}
\end{figure*}

Figure~\ref{gap2-g1.7} shows results in the tetragonal phase ($g=-1.7t$). 
The second largest eigenvalue is characterized by $d_{x^2-y^2}$-wave symmetry, which is nearly 
degenerate to the leading $s_{++}$-wave symmetry [see \fig{lambda-T}(a)]. 
The third one is characterized by $s_{\pm}$-wave symmetry, which is the same symmetry as that often obtained in 
a spin fluctuation mechanism \cite{kuroki08,mazin08,chubukov08}. 
The fourth one corresponds to $d_{x^2-y^2}$-wave symmetry. The difference from the second one 
lies in the sign of the gap on FS3 and FS4. The fifth one is characterized by $s_{++}$-wave symmetry, 
the same symmetry as the leading one [\fig{gap}(a)]. The main difference appears 
in the magnitude of the gap on FS3 and FS4, which is substantially suppressed for the 
fifth leading instability. Looking through those gap structure of the subleading instabilities, 
we can conclude that the momentum dependence of the pairing gap is very weak for 
the $s$-wave solutions in the tetragonal phase.

\begin{figure*}[t]
\centering
\includegraphics[width=15cm]{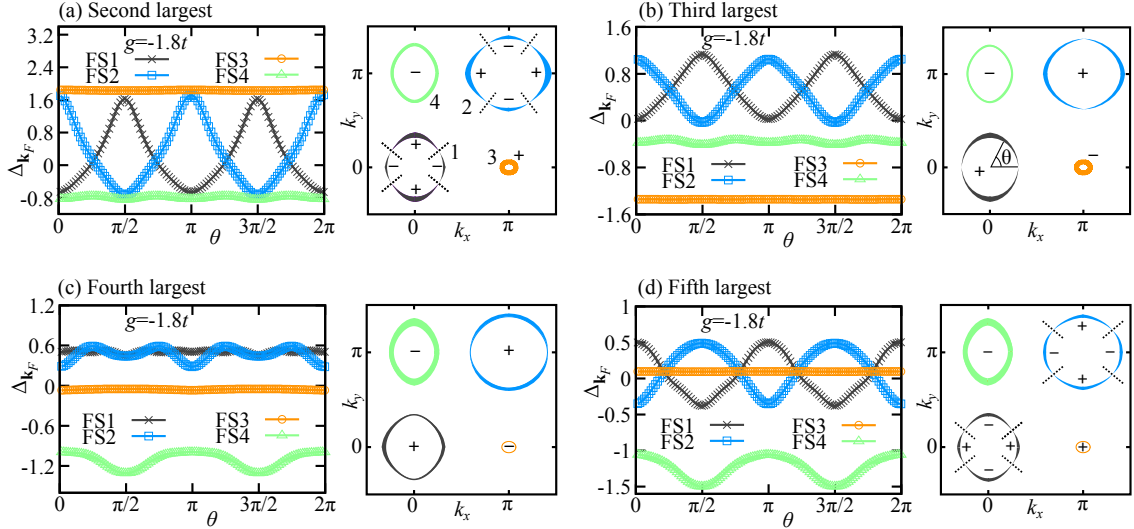}
\caption{(Color online) 
Momentum dependence of the pairing gap for the second (a), third (b), fourth (c), and fifth (d) 
largest eigenvalues in the nematic phase for $g=-1.8t$. 
The right-hand panel is a sketch of the gap structure by  
featuring a gap magnitude with the thickness of each FS. 
}
\label{gap2-g1.8}
\end{figure*}

In the nematic phase the pairing gap acquires a sizable modulation along the FSs. 
Figure~\ref{gap2-g1.8} is the corresponding results in the nematic phase for $g=-1.8t$. 
The pairing gap for the second largest eigenvalue shows a similar 
momentum dependence to that in \fig{gap2-g1.7}(a). However, $d_{x^2-y^2}$-wave symmetry 
cannot be defined in the nematic phase. Instead, the result in \fig{gap2-g1.8}(a) is 
characterized by nodal s-wave symmetry. Nodes enter hole pockets FS1 and FS2. 
The third leading instability corresponds to the so-called $s_{\pm}$-wave symmetry although $s$-wave gap on 
FS1 and FS2 becomes nearly zero at $\theta =0, \pi$ and $\theta =\pi/2, 3\pi/2$, respectively. 
It is interesting that the pairing gap on FS3 becomes largest 
in Figs~\ref{gap2-g1.8}(a) and (b), although FS3 is tiny. 
The fourth one is also characterized by $s_{\pm}$-wave symmetry. 
In contrast to the third one,  the gap on FS3 is nearly zero. 
While this is not a leading instability, the solution in \fig{gap2-g1.8}(c)  provides an interesting example 
of essentially gapless $s$-wave SC in a multipocket system. 
The fifth solution [\fig{gap2-g1.8}(d)] is similar to the fourth one in the tetragonal phase [\fig{gap2-g1.7}(c)] 
and features a kind of $d_{x^2-y^2}$-wave symmetry, although the correct symmetry is the so-called $s$-wave 
symmetry classified by $A_1$ representation in the $C_{2v}$ point group. 
The gap on FS3 is nearly zero, similar to the case of the fourth solution [\fig{gap2-g1.8}(c)]. 

\begin{figure*}[ht]
\centering
\includegraphics[width=15cm]{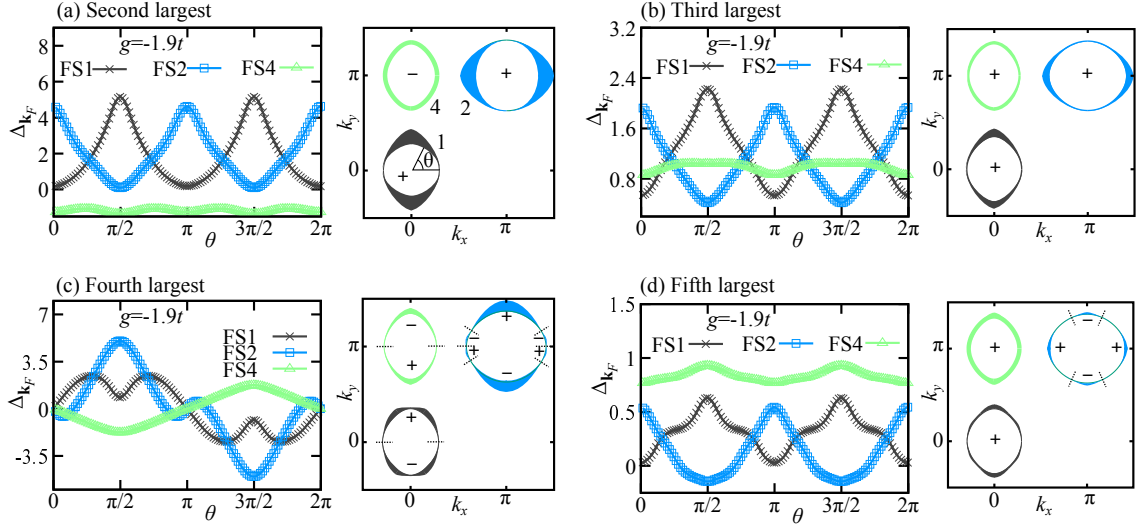}
\caption{(Color online) 
Momentum dependence of the pairing gap for the second (a), third (b), fourth (c), and fifth (d) 
largest eigenvalues deeply inside the nematic phase for $g=-1.9t$ where FS3 vanishes.  
The right-hand panel is a sketch of the gap structure by  
featuring a gap magnitude with the thickness of each FS.  
}
\label{gap2-g1.9}
\end{figure*}

While the leading instability is characterized by the same $s_{++}$-wave symmetry 
in both Figs.~\ref{gap}(b) and \ref{g1.9}(c), the subleading instabilities deeply inside the nematic phase, 
where FS3 vanishes, exhibit symmetries rather different from \fig{gap2-g1.8}. 
Figure~\ref{gap2-g1.9}(a) shows $\Delta_{\vk_F}$ corresponding to the second largest eigenvalue. 
It is characterized by a very large modulation on FS1 and FS2 and the gap almost vanishes  at 
$\theta=0, \pi$ on FS1 and $\theta=\pi/2, 3\pi/2$ on FS2. This solution is similar to $\Delta_{\vk_F}$ 
of the third largest eigenvalue for $g=-1.8t$ [see \fig{gap2-g1.8}(b)]. 
The gap on FS4 has the sign opposite to that on the hole FSs. 
In this sense the gap structure is $s_{\pm}$-wave symmetry. 
Figure~\ref{gap2-g1.9}(b) corresponds to the third largest eigenvalue and 
is a similar result to \fig{gap2-g1.9}(a), except that the gap on FS4 has the opposite sign and  
the modulation of the gaps on FS1 and FS2 is smaller. In fact, these two solutions are almost degenerate 
as seen in \fig{g1.9}(a). 
The fourth largest eigenvalue corresponds to $p$-wave symmetry, as shown in \fig{gap2-g1.9}(c). 
It is interesting to recognize that a $p$-wave solution, in principle, can be driven 
orbital nematic fluctuations deeply inside the nematic phase. 
This $p$-wave solution is almost degenerate to the fifth leading instability as seen in \fig{g1.9}(a). 
The fifth one is nodal $s$-wave symmetry with nodes on FS2. 
A node-like feature is also realized on FS1 where 
the gap nearly vanishes at $\theta=0,\pi$.

\bibliography{main} 

\begin{thebibliography}{53}
\bibitem [{mis()}]{misc-g}%
  \BibitemOpen
  \href@noop {} {}\bibinfo {note} {In the two-band Hubbard model the coupling
  strength $g$ is expressed by $g=U/4-U'/2+J/4$, where $U$ and $U'$ are the
  intra- and inter-orbital Coulomb interaction, respectively, and $J$ is the
  Hund coupling.}\BibitemShut {Stop}%
\end{thebibliography}

\end{document}